***************************************************************************

\documentstyle[12pt]{article}
\def\x{{\bf x}}

\def\k{{\bf k}}

\def\u{{\bf u}}

\def\x{{\bf x}}

\def\begineq{\begin{equation}}
\def\endeq{\end{equation}}
\begin{document}
\title{Universality in fully developed turbulence}
\author{Siegfried Grossmann $^1$ and Detlef Lohse $^{1,2}$}
\maketitle

\bigskip

\begin{tabular}{ll}

$^1$ & Fachbereich Physik, Philipps-Universit\"at,\\
& Renthof 6, D-35032 Marburg, Germany \\\\
$^2$ & The James Franck Institute, The University of Chicago,\\
& 5640 South Ellis Avenue, Chicago, IL 60637, USA\\\\
\end{tabular}

\date{}
\maketitle
\bigskip
\bigskip

We extend the numerical    simulations    of    She    et    al.
[Phys.\ Rev.\ Lett.\ 70, 3251 (1993)]
of highly turbulent flow with $15
\le$  Taylor-Reynolds number $Re_\lambda\le 200$ up
to  $Re_\lambda \approx 45~000$,  employing a reduced wave  vector  set
method  (introduced earlier) to approximately solve  the
Navier-Stokes
equation.   First, also  for  these  extremely  high   Reynolds numbers
$Re_\lambda$, the energy spectra as well as the higher moments
-- when scaled by the spectral intensity at
the wave number $k_p$ of peak dissipation --
can be described by
{\it one universal} function
of $k/k_p$
for all $Re_\lambda$.
Second, the ISR scaling  exponents
$\zeta_m$ of this universal function are
in agreement with the 1941 Kolmogorov theory (the better, the large
$Re_\lambda$ is),
as is the $Re_\lambda$ dependence
of $k_p$. Only
around $k_p$ viscous damping leads to slight energy pileup in the spectra,
as in the experimental data (bottleneck phenomenon).



\newpage
\section{Introduction}
\subsection{Universal turbulent spectra}
Very high Reynolds number turbulence  still resists  full  numerical
simulations.  While in experiments Taylor-Reynolds
numbers $Re_\lambda$ up to $Re_\lambda = 13~000$ have been reported
\cite{Gra61}, the most turbulent {\it numerical} flow
has -- to our knowledge -- been realized by She et al.  \cite{she93},
who achieve $Re_\lambda =200$ at a resolution of $512^{3}$.  These authors
find that  for   all   $Re_\lambda$   up   to
$Re_\lambda =200$  all their energy  spectra  coincide   when  scaled
by  the   spectral intensity at
the
wave number  $k_p$  of  peak  dissipation. I.e., the function
\begineq
\langle |\u(\k) |^{(2)}\rangle /
\langle |\u(\k_p)|^{2}\rangle = F^{(2)} (k / k_p),
\label{eq1}
\endeq
is  universal as assumed  by  Kolmogorov and Obukhov
\cite{Kol41} -- both in the
inertial subrange (ISR) and in the viscous subrange
(VSR).  Note that $F^{(2)}(1)=1$ by definition. The
universality is also found  in experimental
spectra \cite{she93b}. For further numerical simulations, see also
\cite{ker85}.

Kolmogorov and Obukhov \cite{Kol41} not only  assumed
universality of $F^{(2)} (k/ k_p)$,  but also its power law behavior
$F^{(2)}(k/k_p) \propto    (k/k_p)^{-\zeta_2}$ in the ISR with
the classical exponent
$\zeta_2=2/3$.
(The scaling exponent $2/3$ in the discrete Fourier transform,
which we use here,
corresponds to $5/3$ in the continuous case.)

It has been argued in a long lasting debate
(see e.g.\ \cite{kol62,ans84,men91b}) that there are small
intermittency  corrections  to  the  classical  scaling  exponent
$\zeta_2=2/3$.   Unfortunately,  even  at  today's  state  of  the
computational art \cite{she93},  $\zeta_2$ cannot sufficiently precisely
be  determined
from  full numerical  simulations
so that one could confirm or rule out deviations from 2/3. The reason is
that  for the tractable $Re_\lambda$
the  available wave number range   is  quite
narrow ($ k_p/k_{min} \approx 5$ in   \cite{she93};
$k_{min}$ is the lowest wave number free of forcing).
To  obtain an ISR which extends over  more  than  a
decade, $k_{p}$ must be larger   than $50  k_{min}$ \cite{she93}.
To   realize  this,   a  resolution  $\ge  1500^{3}$  is   required
\cite{she93}. The
required      computer      work      increases      as
$Re_\lambda^{6}\log_2{Re_\lambda}$ \cite{orz70}.  We are thus far  away
from  being  able to create developed
turbulence in a  numerical  flow  for as
high $Re_\lambda$ as in experiment \cite{Gra61}.
The huge gap between experiments and simulations is demonstrated in
table 1.
We therefore still need reasonable
approximation techniques to  numerically
solve the Navier-Stokes equation.

\subsection{Reduced wave vector set approximation for high $Re$-turbulence}
Such an approximation has been introduced by us
in  \cite{gnlo92b,egg91a}.  Meanwhile we could  considerably
improve our approach \cite{gnlo93f}.  Here we employ  it
to study universal features of fully developed turbulence  for
Reynolds  numbers $Re$ between $730$ and $1.4\cdot 10^{7}$
(corresponding to Taylor-Reynolds
numbers  $Re_\lambda$ between $120$ and $45~000$, see table 2 below).
This leads to remarkably large ISRs.
For $Re=1.4\cdot 10^{7}$,     the     extension     of    the     ISR     is
$k_p/k_{min}\approx 2000$,  i.e.,  more than three decades, compared
to  $k_p/k_{min}\approx 5$  achieved  in  the  presently best
Navier-Stokes
simulation \cite{she93}.

For completeness we shortly repeat the main idea of our approximation.
To deal feasibly with the  many  scales
present  in  turbulent flow,  we only  admit  a  geometrically
scaling   subset   $K$  of  wave  vectors  in  the   Fourier   sum,
$K=\cup_l K_l$,      thus
$u_i(\x,t)=\sum_{\k\in K}  u_i(\k,t) \exp{(i\k\cdot\x )}$.
$K_0=\{\k_n^{(0)}$,  $n=1,...,n_{max}=80\}$ contains
appropriately chosen wave vectors,
$K_l=\{\k_n^{(l)}=2^l \k_n^{(0)},
\quad n=1,...,n_{max}=80\}$, $l=1,...,l_{max}$,
are scaled replica of $K_0$.
The choice of $l_{max}$ depends on the control parameter  $\nu$,
the viscosity.
The incompressible Navier-Stokes  equation  is
solved on $K$
with periodic boundary conditions in a box of size $(2\pi L)^3$.
All lengths will from now on be measured in multiples of $L$, so
the smallest component of the smallest wave vector is $1$.
The flow is permanently, non-stochastically forced on the outer length
scale with energy input rate $\epsilon$.
All  times will henceforth
be given in multiples of $(L^2/\epsilon)^{1/3}$, i.e.,
the energy
input and thus in the stationary case also the energy dissipation rate
is $\epsilon=1$.
The  type  of  forcing does  not  influence  our  results
sizeably \cite{gnlo93f}.
The smallest wave vectors whose amplitudes are free of forcing have
length $k_{min}=3$.
The smallest wave vectors
at all are $\pm (1,1,-2)$ + permutations and thus
have length $\sqrt{6} = 0.82 k_{min}$. All wave vector
amplitudes $\u (\k ,t)$
with $|\k | \ge k_{min}=3$ are free of forcing.

\subsection{Contents of the paper}
The paper is organized as follows: In section 2 we examine the
$Re$-dependence of the Taylor-Reynolds number. In section 3 we
confirm universality up to the highest $Re$ we can treat. The
form of the universal spectra is discussed in section 4.
Section 5 is left for a short summary of our findings.

\section{$Re$-dependences}
We now come to our results. First, what are  the Reynolds numbers $Re$ of
our approximate numerical turbulence?
As usual, there is some arbitrariness in the definition of the
Reynolds number. We regard $L_0=\lambda/2$ as the outer length scale,
where $\lambda$ is the wave length of the smallest wave vector, and
$U_0=2u_{1,rms}$ as the
typical velocity difference on the outer lengthscale. Thus
$Re = U_0L_0/\nu$ can be considered as an
appropriate definition of the Reynolds number. The data for five
simulations, covering four decades of $Re$, are given in table 2.
We also list  the Taylor Reynolds numbers $Re_\lambda=u_{1,rms}
\lambda_{Taylor}/\nu$, where
$\lambda_{Taylor}=u_{1,rms}/ (\partial_1u_1)_{rms}$ are
the Taylor lengths.

The dissipation rate is balanced by the input rate $\epsilon \sim
U^3_0/L_0$ \cite{ll87,my75}. We therefore write
\begineq
\epsilon=c_\epsilon U^3_0/L_0,
\label{eq_eps}
\endeq
where $c_\epsilon$ is a dimensionless number.
Since we choose $\epsilon =1$, $L_0=\pi/\sqrt{6}$, and since we find
$U_0 =2u_{1,rms}$ from the numerical solution, we can determine $c_\epsilon$
from this equation. It turns out to decrease with increasing $Re$,
seemingly to a final level somewhere near $6\cdot 10^{-3}$.
Note that
for laminar flow, on the other hand,  it holds
$c_\epsilon \propto Re^{-1}$, see e.g.\ \cite{ll87}.

Equation (\ref{eq_eps})  leads to the relation
\begineq
Re_\lambda =\sqrt{{15\over 16 c_\epsilon}}\sqrt{Re}.
\label{eq_rel}
\endeq
When $c_\epsilon$ eventually becomes universal, i.e.,
independent of $Re$,  the well known
large $Re$ limit relation $Re_\lambda\propto \sqrt{Re}$ is recovered.
In experiment, for smaller $Re$ the measured Taylor Reynolds  number
$Re_\lambda$ turns out to be
smaller than predicted by $Re_\lambda\propto \sqrt{Re}$
\cite{cas93b,lat92}, as in fact $c_\epsilon$ depends
on $Re$. Both the large
$Re$ behaviour of $Re_\lambda$
with the power law exponent $1/2$ and the deviations for smaller $Re$
can also be
seen in our approximate solutions, see Fig.\ \ref{f_rel}.
For a more detailed discussion on the $Re$-dependence of $c_\epsilon$
and $Re_\lambda$, see \cite{loh94,sre84}.

One remark concerning the nominal value
of the prefactor in eq.\ (\ref{eq_rel}).
Taking the large Reynolds numbers, we have
(with $c_\epsilon =6.5 \cdot 10^{-3}$
from table 2) $Re_\lambda \approx 12\sqrt{Re}$,
whereas experimentally it is
$Re_\lambda \sim \sqrt{Re}$, i.e., we overestimate the Taylor Reynolds numbers
by one order of magnitude. We explain this as
due to our approximation, as in our reduced wave vector set $K$ the
larger wave vectors
are considerably thinned out. So less energy than ought to  be
is transported downscale, leading to a larger $u_{1,rms}$ than in real
turbulence and thus to a larger $Re_\lambda = u_{1,rms}^2/\nu
(\partial_1
u_1)_{rms}$. The same blocking
effect by phase space sparseness at larger $k$
also leads to an overestimation of the Kolmogorov
constant $b$ in the velocity structure function $D(r)=b(\epsilon r)^{2/3}$
as already mentioned  in \cite{gnlo93f,egg91a} -- also by a factor of about
10 ($b\approx 70$ instead of $b=8.4$ as in experiment, see \cite{gnlo93f}),
because both $Re_\lambda$ and $D(r)$ are quadratic in $u_{1,rms}$.
(Note that the denominator $(\partial_1 u_1)_{rms}$ is proportional
to $\sqrt{\epsilon}$ and thus kept fixed.)

\section{Universal wave number spectra}
The  wave number spectra of our approximate solutions are  shown  in  Fig.
\ref{f_unis}.
We can confirm that
$F^{(2)}(k/k_p)$   in   fact is  {\it universal}
for   all   $Re$ even much beyond
$Re_\lambda =200$
as studied in \cite{she93}.
As   in
\cite{she93},  the wave number $k_p$ of peak dissipation
is found to increase as $Re^{3/4}$, i.e.,
$\eta k_p $  is constant also for the huge $Re$  we
simulated, see table 3. We find $k_p \approx 1/10\eta$.
For  $Re=1.4\cdot 10^{7}$  the ISR extends as
far as 3.3 decades to the  left  of $k_p$, see table 2.
In k-space $\log_{10}
(k_p/k_{min})$ gives the extension of the ISR. This extension of the scaling
range should also follow from the Reynolds number $Re$. In r-space the
scaling occurs
between the outer length scale $L_0$ and $10\eta$, where
$\eta=\nu^{3/4}/\epsilon^{1/4}$
is the Kolmogorov length scale \cite{ll87,my75}. For large $Re$ it is
\cite{ll87}
$$
\qquad\qquad\qquad\qquad\qquad\qquad
L_0/10\eta = c_\eta^\prime Re^{3/4}.
\qquad\qquad\qquad\qquad\qquad\qquad
(4')
$$
Formally eq.\ ($4'$) can be derived from eq.\ (\ref{eq_eps})
and one gets $c_\eta^\prime = c_\epsilon^{1/4}/10$.
Note that thus $c_\eta^\prime$
is also slightly $Re$-dependent via $c_\epsilon (Re)$.
Yet our $c_\epsilon$
is smaller than the experimental one, see last paragraph of section 2.
Therefore
we rather define a $c_\eta$ from the extension of the scaling
range found in our numerical simulation than from ($4'$),
namely by
\begineq
k_{min} /k_{p} = c_\eta Re^{3/4}.
\label{eq_l0}
\endeq
This corresponds to ($4'$) since $L_0$ is approximately
$1/k_{min}$ and $10\eta$ is approximately $1/k_p$. More
precisely we have $L_0 = 3\pi/(\sqrt{6} k_{min})$, so $L_0^{-1} =
0.26 k_{min}$.  The ratio $k_p/k_{min}$ can
be  extracted  from our  numerical spectra in k-space.
For the largest $Re=1.4\cdot 10^7$ we find $k_p/k_{min}=1950$ and thus
obtain
$c_\eta = (k_p/k_{min}) Re^{-3/4}=8.52\cdot 10^{-3}$.
We now disregard the small $Re$-dependence of $c_\eta$ and calculate the
extension of
the scaling ranges for smaller $Re$ from (\ref{eq_l0})
with $c_\eta = 8.52\cdot 10^{-3}$. The results are
given in the last column of table 2 and are found in
excellent agreement with the length of the scaling ranges seen
in our approximate solutions for these smaller $Re$, see 6th column
of table 2. Only for the smallest
$Re$ we find disagreement, as expected, because for $Re=730$ the
constant $c_\eta$ in eq.\ (\ref{eq_l0}) can no longer be considered as
independent of $Re$.

Moreover, also
the $F^{(m)}(k/k_p)=\langle |\u(\k)|^{m}\rangle /\langle
|\u(\k_p)|^{m}\rangle $,
m=3,4,6,8,10,   are found  to  be  {\it universal}   (see Fig.\
\ref{f_unis} for $m=6$).
Due to the very extended ISR we can
determine  the  power law exponents $\zeta_m$ of the  universal   functions
$F^{(m)}( k/k_p)\propto (k/k_p)^{-\zeta_m}$ rather precise,
cf.\ table 3.
This is still  not possible
in full simulations, as the universal scaling range
is too small, $k_p/k_{min} \le 5$ \cite{ker85,she93}. If
a power law fit for full simulations is tried
nevertheless, one gets scaling exponents $\zeta_m$ much smaller than the
classical ones, due to the nonuniversal large scale forcing \cite{ker93}.

\section{Form of the universal spectra}
Having shown  universality, we now check several fits to the
universal spectra $F^{(m)}(k/k_p)$ that we obtained from our numerical data.

\subsection{Power law fit and intermittency}
Firstly,
we fitted the spectra for all $Re$ by the two parameter  function
\begineq
F^{(m)}(\tilde k) = \tilde k^{-\zeta_m} \exp{[- (\tilde k-1 )/\tilde
k_d^{(m)}]},
\qquad \tilde k = k/k_p,
\qquad \tilde k_d^{(m)} = k_d^{(m)}/k_p.
\label{eq2}
\endeq
This form has theoretical support \cite{foi90} and was also successfully used
to fit experimental spectra \cite{sre85,pro91}.
The  crossover
between the power law behaviour in the ISR and exponential fall off in the VSR
takes   place at the wave number
$k_d^{(m)}$.
In table 3 we list the parameters, obtained from a fit in   the   range
$0\le k\le \eta^{-1}/4$ or $0\le k/k_p \le 2.5$, respectively.
The fit (\ref{eq2}) gives $\zeta_3$ rather near, but not
exactly equal to 1 as it should be according
to Kolmogorov's  structure equation  \cite{my75}.
This tiny deviation is corrected in table 3 by dividing the scaling exponents
obtained from the fit (\ref{eq2}) by $\zeta_3$.
The resulting scaling exponents take their classical
values $\zeta_m = m/3$ with high accuracy, see table 3.

{} From  the appearance of the classical  scaling  exponents
$\zeta_m=m/3$ one might deduce that there is no intermittency  in
our  signal.
This conclusion would not be correct.  In  fact,  for small scales
we  do observe  strong intermittency   in  the  signal
\cite{gnlo92b,gnlo93f}. We therefore suggested to
introduce local $\zeta_m (k)$, defined by local fits of type
(\ref{eq2}) in the restricted k-ranges
$[k/\sqrt{10},  k\sqrt{10}]$ for  all  $k$,  keeping  the
$k_d^{(m)}$ fixed at their global values \cite{gnlo93f}.

The  surprising  result  is  shown  in  Fig.\ \ref{f_sri}.  There  are  large
intermittency  corrections  $\delta\zeta_m(k)= \zeta_m(k)-m/3$  at
small scales (VSR),  only moderate intermittency corrections  at
large scales (stirring subrange,  SSR), but hardly any deviations
from  classical scaling in the ISR.  This astonishing result  was
extensively discussed already in \cite{gnlo93f}. Here it can be
confirmed for a considerably larger $Re$-range.

In addition, we fitted eq.\ (\ref{eq2}) to our spectra, but now
with $k_d^{(m)}$ {\it fixed} at
$k_d^{(m)}=2k_d^{(2)}/m$, $k_d^{(2)}= (13.5\eta )^{-1}$, see below.
Again we find only tiny global intermittency corrections, which clearly
{\it decrease} with increasing $Re$, as predicted by
\cite{cas93b,lvo94}. For $Re= 1.05 \cdot 10^4$ we find
$\delta\zeta_2 = 0.012$, $\delta\zeta_6 = -0.058$, going down to
$\delta\zeta_2 = 0.002$, $\delta\zeta_6 = -0.011$ for
$Re= 1.4 \cdot 10^7$, suggesting, that intermittency might be
a finite size effect. For details and a theoretical explanation
we refer to ref.\ \cite{gllp94}.

We now discuss the crossover scale between ISR and VSR. From the fit
(\ref{eq2}) we get
$(k_d^{(2)})^{-1}\approx 13.5\eta$,
i.e., the crossover scale is
one order of magnitude larger than the Kolmogorov length. This
fact  has long been known from
experiment  \cite{my75}  and theory  \cite{eff87}.  From
the maximum condition for $ k^2 F^{(2)}( k)$ at $ k =k_p$
it
follows that $\tilde k_d^{(2)}=1/(2-\zeta_2)=3/4$.  In our simulations
there are some fluctuations around this value.  The reason is that we
have {\it discrete} wave vectors which are not dense in the
VSR.  Thus  $k_p$ can only be determined with limited  accuracy,
cf.\ Fig.\ \ref{f_unie},
where     the     energy     dissipation      rate
$\epsilon (k)=\k^2 \langle |\u(\k)|^{2} \rangle$ is  displayed.
Typically the relative k-distances are $\delta k/k \approx 1/10$, which
corresponds to the $\approx 10 \%$ deviations of $\tilde k_d^{(2)}$ from
$0.75$ in table 3.
Of course,  to increase the accuracy of $k_p$,  one could also define
it  in  terms  of $k_d^{(2)}$ from the global fit  (\ref{eq2})  with  fixed
$\zeta_2=2/3$ to be $k_p=4k_d^{(2)}/3$.
The resulting small changes are not visible in Fig.\ \ref{f_unis}.

Where does the crossover from ISR to VSR behaviour take place in
higher    order  moments?     From     the    fit    (\ref{eq2})
we     find
$k_d^{(m)}=2 k_d^{(2)}/m$, i.e.,
$\tilde k_d^{(m)}=3/(2m)$  to a very  high
precision \cite{gnlo93f}.  This  means  that  the  ISR  is
considerably smaller for higher order moments, or, to state it differently,
higher order moments feel viscosity earlier than lower order moments.

\subsection{Energy pileup in the crossover region between ISR and VSR}
We also applied fits different from (\ref{eq2}), as
from Fig.\ \ref{f_she}
it might seem that (\ref{eq2}) only badly fits the spectrum  in
the range around $k_p$. The same observation was reported already
by She and
Jackson  \cite{she93b}  when they determined  the  universal  function
$F^{(2)}(k/k_p)$ from experimental data. To improve the fit, they
suggested to use the empirical three parameter function
\begineq
F^{(2)}({\tilde k }) = {\tilde k^{-2/3} \over 1+\alpha}
\left( 1 + \alpha \tilde k^{\beta }\right)
\exp{[- (\tilde k-1)/\tilde k_d^{(2)\prime}]}, \qquad \tilde k=k/k_p,
\qquad \tilde k_d^{(2)\prime} = k_d^{(2)\prime}/k_p,
\label{eq3}
\endeq
and found $\alpha=0.8$,  $\beta\approx 0.7$,  i.e., near $k=k_p$ the decay
of the spectral power is diminished. Their physical interpretation is
a  pileup of excitation around $k_p$,  possibly due  to  coherent
vortex structures.

We also tried the fit (\ref{eq3})
and found $\alpha\approx 2$,
$\beta\approx 1.8$,
and $  \tilde k_d^{(2)\prime} \approx 0.4$, all
slightly depending on  $Re$.
Thus the energy pileup at $k\approx k_p$
in our approximate Navier-Stokes solution seems to be even
stronger than in experiment, as we have an
additive  correction term
with a  {\it larger} exponent.
We ascribe
the larger correction summand to our approximation: larger wave
vectors are more
and more sparse, so the dissipative scales cannot acquire the kinetic
energy as fast as they should and pile it up around $k_p$. Of course,
$F^{(2)}(k/k_p)$ does not increase with $k/k_p$ as the correction term is
strongly damped by $\exp{(- k/k_d^{(2)\prime})}$
with $k_d^{(2)\prime}$
much smaller as $k_d^{(2)}$
before.
The energy pileup can also be observed in Fig.\ \ref{f_she}, where we replotted
$F^{(2)} (k/k_p) $ with the two fits eq.\ (\ref{eq2}), where we fixed
$\zeta_2=2/3$, and eq.\ (\ref{eq3}).
For $k\approx k_p$ the
fit (\ref{eq3}) is slightly superior to the fit (\ref{eq2}).
We have to raise doubts that
the energy pileup is due to coherent vortex structures, as was speculated in
\cite{she93b}, because these are less in our approximation than in experiment,
cf.\
\cite{gnlo92b}, while our pileup is stronger than found in the measured data.

Instead,   the energy pileup may possibly be explained by  the socalled
bottleneck phenomenon \cite{fal94}. This phenomenon can be described
as follows: Imagine a triad Navier-Stokes  interaction between the amplitudes
$\u (\k_1)$, $\u(\k_2)$, and $\u(\k_3)$, $\k_1+\k_2+\k_3=0$, $k_1 < k_{2,3}$,
so that $\u(\k_2)$ and $\u(\k_3)$ are already considerably damped by
viscosity,
in additon to the power law decrease $\propto k^{-2/3}$.
So the turbulent energy transfer
downscale $\sim k_1 u(k_1)u(k_2)u(k_3)$ would be
reduced and stationarity could not be achieved, if $\u(\k_1)$ did not
increase, i.e., an energy pileup at  $k_1$ is established.
The effect is strongest if $k_1$ is around $k_p$, because there
$\u(\k_2)$ and $\u(\k_3)$ are already considerably damped.
Of course, there is also viscous
damping around $k_1$, which would counteract the bottleneck effect,
but for $k_1 < \eta^{-1}$ the damping by viscosity $\nu$  is weaker than the
damping by the {\it eddy} viscosity \cite{fal94}.

For $k \ll k_p$ Falkovich \cite{fal94} calculates the first order correction
to the K41 spectrum due to the bottleneck phenomenon and obtains
\begineq
F^{(2)} (\tilde k ) \propto
F_{K41}^{(2)} (\tilde k )  [1 + \tilde k^{4/3}/ \log(\tilde k^{-1})],
\qquad \tilde k = k/k_p.
\label{eqfal}
\endeq
We tried to  compare our results
for the scaling of the correction term
with his prediction $\propto \tilde k^{\beta^\prime}$, $\beta^\prime =4/3$
(apart from the log-correction), but as $k \ll k_p$
(i.e., $\tilde k \ll 1$)  has to hold, we can
only fit with a very limited number of data points. As in addition
for small $k$ the design matrix of the fitting problem nearly
degenerates, our results strongly depend on the fit range we choose and
we can neither verify nor rule out his prediction $\beta^\prime = 4/3$.

The same pileup as for $F^{(2)}(k)$ also
appears in higher order velocity moments $\langle |\u(\k)|^m
\rangle$ with $m>2$, leading to smaller local scaling exponents $\zeta_m(k)$
for $k$ near the VSR. Possibly this effect mimicks intermittency
corrections to classical scaling
in experimental data or simulated data with shorter scaling ranges
which in fact would not show up in sufficiently long
ISR for large enough $Re$.
This interpretation is also supported by the behavior
of the scale resolved
intermittency exponents, $\zeta_m(k)$, see Fig.\ \ref{f_sri} above and
ref.\ \cite{gnlo93f}.

\subsection{Log-similarity description of the spectra}
Finally,
besides the normalization of the spectra
(\ref{eq1}) another procedure has been suggested
to get an universal description of the experimental data,
namely, the log-similarity
description \cite{cas93b}. This  claims  that the
logarithmic spectra coincide for different $Re_\lambda$, when both the
abszissa  and the ordinate are multiplied by some function $\beta (Re_\lambda
)$, i.e., $\beta \log (|\u (\k )|^2/|\u(\k_0)|^2)$ against
$\beta \log (k/k_0)$ is claimed to be
universal. $k_0$ is a wave number which has to be fitted
to the experimental data.
In \cite{cas93b} $k_0 \approx 2k_p \approx (6\eta )^{-1}$
is found. For large $Re_\lambda$ it is $\beta = 0.9/ \log (Re_\lambda / 75 )$
\cite{cas93b}. We plot the spectra for our three largest $Re$ in this
parametrization, see inset of Fig.\ \ref{f_she}.
As Taylor Reynolds number we simply take
$\sqrt{Re}$, because our approximation overestimates $Re_\lambda$, see the
discussion in section 2.
For smaller $Re$ the function $\beta (Re_\lambda )$
behaves quite different, so it is not reasonable to show also  the
spectra for smaller $Re$ in the
plot. The quality of the superposition of the spectra might improve  if one
readjusted the free parameters of this description.

\section{Summary}
To summarize,  we first demonstrated the {\it universality} of the normalized
moments  $F^{(m)}(k/k_p)$  up  to  $Re=1.4\cdot 10^{7}$  within  our
approximate Navier-Stokes solution.
Second, the $F^{(m)}( k/k_p)$ can pretty well be described by
\begineq
F^{(m)}(\tilde k) = \tilde k^{-m/ 3}
\exp{ \left(-{2m\over 3}  (\tilde k-1) \right)}, \qquad \tilde k=k/k_p .
\label{eq4}
\endeq
Near peak dissipation
an additive correction to the classical
scaling $F^{(m)}(k/k_p)\propto (k/k_p)^{-m/3}$ shows up, which might be
explained by  an energy pileup
around $k_p$ due to a viscosity induced bottleneck phenomenon \cite{fal94}.
This bottleneck phenomenon  might mimick
intermittency corrections to classical scaling in experimental or
simulated data with less resolution and for smaller Reynolds numbers.

\newpage
\noindent
{\bf Acknowledgements:}
D.\ L.\ thanks B.\ Castaing, G.\ Falkovich,
Th.\ Gebhardt, and L.\ Kadanoff for very helpful discussions and
the Aspen Center of Physics  for its hospitality.
Partial support by the German-Israel-Foundation (GIF), by DOE, and by
a NATO grant, attributed by the Deutsche Akademische Austauschdienst (DAAD),
is gratefully acknowledged.
The   HLRZ  J\"ulich
supplied us with computer time.

\newpage

\centerline{\bf Tables}

 \begin{table}[htp]
 \begin{center}
 \begin{tabular}{|r|r|r|r|r|}
 \hline
         $   $
       & $Re$
       & $Re_\lambda$
       & $k_p/k_{min}$
       & $\#$ of modes
 \\
 \hline
         best experiments \cite{Gra61}
       & $1.7 \cdot 10^8$
       & $13~000$
       & $\approx 10~000$
       & $\infty$
 \\
         best simulations  \cite{she93}
       & $\approx 4\cdot 10^4 $
       & $200$
       & $5$
       & $4\cdot 10^8$
 \\
         REWA [here]
       & $1.4 \cdot 10^7$
       & $45~000$
       & $\approx 2~000$
       & $80\cdot 3 \cdot 13 = 3120$
 \\
 \hline
 \end{tabular}
 \end{center}
 \end{table}

\vspace{1cm}

\centerline{\bf Table 1}
In the first two lines $Re$, $Re_\lambda$, the length of the scaling
range $k_p/k_{min}$, and the number of contributing modes are compared
for the most developed experimental and numerical turbulence,
respectively. In the third line we give the same data for
the largest $Re$ of our
reduced wave vector set approximation (REWA).

\vspace{2cm}

 \begin{table}[htp]
 \begin{center}
 \begin{tabular}{|r|r|r|r|r|r|r|}
 \hline
         $\nu$
       & $l_{max}+1$
       & $Re$
       & $Re_\lambda$
       & $c_\epsilon(Re)$
       & $\log_{10}(k_p /k_{min})$
       & $\log_{10}(c_\eta Re^{3/4})$
 \\
 \hline
         $5\cdot 10^{-3}$
       & $5$
       & $730$
       & $122$
       & $46\cdot 10^{-3}$
       & $0.24$
       & $0.08$
 \\
         $5\cdot 10^{-4}$
       & $7$
       & $1.05\cdot 10^4$
       & $801$
       & $15\cdot 10^{-3}$
       & $0.97$
       & $0.95$
 \\
         $5\cdot 10^{-5}$
       & $9$
       & $1.25\cdot 10^5$
       & $3590$
       & $9.1\cdot 10^{-3}$
       & $1.79$
       & $1.75$
 \\
         $5\cdot 10^{-6}$
       & $11$
       & $1.37\cdot 10^6$
       & $13\ 600$
       & $7.0\cdot 10^{-3}$
       & $2.48$
       & $2.53$
 \\
         $5\cdot 10^{-7}$
       & $13$
       & $1.40\cdot 10^7$
       & $44\ 800$
       & $6.5\cdot 10^{-3}$
       & $3.29$
       & $3.29$
 \\
  \hline
 \end{tabular}
 \end{center}
 \end{table}

\vspace{1cm}

\centerline{\bf Table 2}
Results from our approximate solutions of the
Navier-Stokes  equation for various $\nu$.
$l_{max} +1$ is the number of rescaled wave number replica $K_l$.
The definition of $Re$ is $Re=U_0 L_0/\nu$.
Here $L_0=\pi/\sqrt{6}$, and $U_0 = 2 u_{1,rms}$
describes the velocity difference across the outer scale,
being determined for each
$\nu$ from our numerical solution.
($u_{1,rms}$  depends on $\nu$
and increases from 1.42 to 2.73 for the $\nu$ in the table.)
$Re_\lambda =u_{1,rms} \lambda_{Taylor} /\nu$,
as usual.
The coefficient
$c_\epsilon$ is calculated according to eq.\ (\ref{eq_eps}).
In the last two columns the extension of the scaling range,
$\log_{10}(k_p /k_{min})$ (found from  the numerical solution),
is compared with that calculated from
eq.\ (\ref{eq_l0}), see text.

\newpage

\vspace{1.5cm}

 \begin{table}[htp]
 \begin{center}
 \begin{tabular}{|r|r|r|r|r|r|}
 \hline
          $\nu$
       & $\zeta_2$
       & $\zeta_6$
       & $(k_p\eta )^{-1}$
       & $(k_d^{(2)}\eta)^{-1}$
       & $\tilde k_d^{(2)}$
 \\
 \hline
         $5\cdot 10^{-3}$
       & $0.656$
       & $1.997$
       & $10.2$
       & $11.8$
       & $0.87$
 \\
         $5\cdot 10^{-4}$
       & $0.678$
       & $1.986$
       & $10.6$
       & $13.6$
       & $0.78$
 \\
         $5\cdot 10^{-5}$
       & $0.672$
       & $1.975$
       & $9.2$
       & $13.7$
       & $0.67$
 \\
         $5\cdot 10^{-6}$
       & $0.669$
       & $1.994$
       & $10.5$
       & $13.3$
       & $0.79$
 \\
         $5\cdot 10^{-7}$
       & $0.668$
       & $1.990$
       & $9.1$
       & $13.3$
       & $0.67$
 \\
  \hline
 \end{tabular}
 \end{center}
 \end{table}

\vspace{1cm}
\centerline{\bf Table 3}
The fit parameters to our approximate solutions of the
Navier-Stokes  equation for the same $\nu$ as in table 2.
$k_p$  is the  wave  number  with  peak
dissipation,  $k_d^{(2)}$ the cut off from our fit (\ref{eq2}), if we fit
the    spectrum   in   the   interval    $[0,\eta^{-1}/4]$. In the last
column their ratio $\tilde k_d^{(2)}=k_d^{(2)}/k_p$ is given.
{} From the condition $\tilde k^2 F^{(2)} (\tilde k) $ maximal at 1, the cut
off should be $\tilde k_d^{(2)}=3/4$.

\newpage

\centerline{\bf Figures}
\begin{figure}[htb]
\caption[]{$Re_\lambda$ (squares) and $c_\epsilon$ (circles) as
functions of $Re$.
Left ordinate is $Re_\lambda$, right ordinate is $c_\epsilon$.
For large
$Re$ the power law $Re_\lambda \propto \sqrt{Re}$ holds, for small $Re$ there
are deviations, as the dimensionless number $c_\epsilon$ in eq.\ (\ref{eq_eps})
depends on $Re$. The dependence of $c_\epsilon$ on $Re$ strongly resembles
the experimental one, cf.\ \cite{cas93b,sre84,ll87} }
\label{f_rel}
\end{figure}

\begin{figure}[htb]
\caption[]{Universal spectra $F^{(m)}(k/k_p)$ for $m=2$ (flatter) and $m=6$
       (steeper)     for     $Re=7.3\cdot10^{2}$ (triangles),
      $Re=1.05\cdot10^{4}$ (crosses),
       $Re=1.25\cdot 10^{5}$ (squares), $Re=1.37\cdot 10^{6}$ (pluses),
     $Re=1.40\cdot 10^{7}$ (diamonds). The arrows indicate the smallest
wave number  {\it free of forcing}, $k_{min}$, for the
    respective $Re$. The smallest wave number {\it of all} $\k \in K$ is
$0.82 k_{min}$, see section 1.2.
The dashed arrow labels $k_{min}$ in the
simulation by She et al.\ \cite{she93}, the dotted arrow marks
$k_{min}$ in the one of
Vincent and Meneguzzi \cite{ker85}.}
\label{f_unis}
\end{figure}

\begin{figure}[htb]
\caption[]{Scale resolved intermittency corrections $-\delta\zeta_m(k)$ for
$m=2,4,6,8,10$, bottom to top. $\nu=5\cdot 10^{-7}$, $Re= 1.4 \cdot 10^7$.
The fit range is
$[ k/\sqrt{10}, k\sqrt{10} ]$ for all $k$.}
\label{f_sri}
\end{figure}

\begin{figure}[htb]
\caption[]{Universal         energy         dissipation          rate
       $\epsilon(k)/\epsilon(k_p)$
 versus $k/k_p$ for $\nu = 5 \cdot 10^{-7}$,
       $Re=1.4\cdot 10^7$.  Also shown
       are the fits resulting from (\ref{eq2}) (dashed) and (\ref{eq3})
(solid).}
\label{f_unie}
\end{figure}

\begin{figure}[htb]
\caption[]{Universal    spectrum    $F^{(2)}(k/k_p)$ for $\nu=5\cdot 10^{-7}$,
 $Re=1.4\cdot 10^7$.
The fits (\ref{eq2}) (dashed) and (\ref{eq3}) (solid) are compared, both with
the same $\zeta_2=2/3$.
Inset: Quality of universality in the
log-similarity description for the second moments, see text. The symbols mean
 $Re=1.25\cdot 10^{5}$ (squares), $Re=1.37\cdot 10^{6}$ (pluses),
     $Re=1.40\cdot 10^{7}$ (diamonds). On the abscissa we plotted
$\beta \log_{10}(k/(2k_p))$, on the ordinate
$\beta \log_{10}(\langle|\u(k)|^2\rangle /
\langle|\u(2k_p)|^2\rangle)$ with $\beta = 0.9/\log(Re_\lambda /75)$,
cf.\ \cite{cas93b}.
}
\label{f_she}
\end{figure}


\newpage


\end{document}